\documentclass[preprint,showpacs,preprintnumbers,amsmath,amssymb,nofootinbib]{revtex4-1}
\usepackage{graphicx}
\usepackage{dcolumn}
\usepackage{amssymb}
\usepackage{mathrsfs}
\usepackage{amsmath}
\usepackage{epsfig}
\usepackage{hhline}
\usepackage[utf8]{inputenc}
\usepackage{color}
\usepackage{multirow}
\usepackage[T1]{fontenc}
\usepackage{verbatim}
\usepackage{slashed}
\usepackage{epstopdf}
\usepackage{lineno,hyperref}
\usepackage{cancel}
\usepackage{xcolor}
\usepackage{fancyhdr}

\begin{document}

\title{Explanation of nearby SNRs for primary electron excess and proton spectral bump}

\author{Tian-Peng Tang$^{1,2}$}
\author{Zi-Qing Xia$^{1}$}
\author{Zhao-Qiang Shen$^{1}$}
\author{Lei Zu$^{1}$}
\author{Lei Feng$^{1,2,3}$}
\email{Corresponding author: fenglei@pmo.ac.cn}
\author{Qiang Yuan$^{1,2}$}
\author{Yi-Zhong Fan$^{1,2}$}
\author{Jian Wu$^{1,2}$}

\affiliation{
$^1$Key Laboratory of dark Matter and Space Astronomy, Purple Mountain
Observatory, Chinese Academy of Sciences, Nanjing 210033, China \\
$^2$School of Astronomy and Space Science, University of Science and Technology of China, Hefei, Anhui 230026, China \\
$^3$Joint Center for Particle, Nuclear Physics and Cosmology,
Nanjing University -- Purple Mountain Observatory,  Nanjing  210093, China
}

\begin{abstract}
  Several groups have reported a possible excess of primary electrons at high energies with the joint fit of the positron fraction and total electron/positron spectra. With the latest release of high-precision electron/positron spectra measured by AMS-02, we further confirm this excess by fitting $\Delta\Phi$ $\rm(i.e., \Phi_{e^-}-\Phi_{e^+})$ data in this work. Then we investigate the contribution of a single nearby supernova remnant to the primary electron excess and find that Monogem can reasonably account for this excess. Moreover, we predict that the electron spectrum may harden again at a few TeVs due to Vela's contribution. DAMPE, which can accurately measure electrons at TeV scale, is expected to provide the robust test of this new spectral feature in the near future. Finally, we fit the proton spectrum data of DAMPE with Monogem or Loop I. We find that both the primary electron excess and the proton spectral bump could be mainly generated by Monogem.
\end{abstract}
\maketitle

\section{Introduction}
In the history of cosmic ray (CR) measurements, electrons were not detected until the 1960s because of their rarity (i.e., they accounted for about one percent of the total CRs). However, this does not prevent electrons from playing an important role in the study of CRs.  With the development of spectroscopic technology, the measurement of CRs has reached unprecedented accuracy, especially the Alpha Magnetic Spectrometer\footnote{\url{https://ams.nasa.gov/}} (AMS-02)~\cite{ams-02} and Dark Matter Particle Explorer (DAMPE) \cite{dampe,dampe-line}. It has been proved that the observed data of positron fraction and total electron/positron spectra both show the excess relative to the conventional background~\cite{1,2,3,4,5,6,7,8}. More evidence is needed, though, to determine whether these excesses are due to new physics or high energy astrophysics.

A spectral hardening was certified in several CR proton measurements~\cite{atic-proton,cream-proton,pamela-proton,ams02-proton}. Inspired by such phenomenon, a similar feature was speculated in the primary CR electron spectrum~\cite{feng2014}. More interestingly, the joint fitting of AMS-02 positron ratio data and CR electron plus positron data shows that there is obviously tension between these two data sets and it can be eliminated by this primary CR electron spectrum hardening model~\cite{9,10,Lin,li2015}.

What is the origin of the primary CR electron excess? The widely discussed potential electron/positron sources, including pulsars and dark matter (DM) annihilation, can be ruled out. Because some studies have suggested that electrons/positrons were produced symmetrically in pairs in these sources~\cite{Profumo:2008ms,Malyshev:2009tw,Hooper:2009fj,Bergstrom:2009fa}. But for Pulsar Wind Nebulae (PWNe), this needs to be further proved.
Unlike other CRs, electrons, due to their light mass, interact with the interstellar medium during their propagation, producing significant radiation and rapidly losing its energy (i.e., the radiation intensity of a charged particle is inversely proportional to the square of its mass). Thus, for the electron at the energy of about 1 TeV, its cooling time scale is about $10^5$ yr and its propagation distance is about 1 kpc. So the source of such high energy electrons should be not far away from the solar system. For the primary electron excess at the energies up to $\sim 0.4$ TeV, the promising sources are the nearby and middle-age astrophysical sources~\cite{li2015,14,15,Fan:2010yq,16,Fornieri:2019ddi}.

Discrete instantaneous supernova remnants (SNRs), in particular, are leading candidates, whose lifetime is estimated as $\tau \ge 4\times10^5(R/1\, \text{kpc})^2(E_e/100\,\text{GeV})^{-1/3}$~\cite{17,18,19,20}. Too old SNRs are not suitable candidates because the CR electrons they generated have diffused very far (i.e., the flux of CRs $\propto \tau^{-3/2}$) and cooled down significantly. So the distance and age make the number of SNRs that meet these requirements very limited. Monogem ($\tau=1.1\times10^5\,\text{yr}, R=0.29\,\text{kpc}$), Loop I ($\tau=2\times10^5\,\text{yr}, R=0.17\,\text{kpc}$) and Geminga ($\tau=3.4\times10^5\,\text{yr}, R=0.25\,\text{kpc}$) may dominantly contribute to the primary electron excess at hundreds of GeV \cite{21,22,Ding:2020wyk,23}. And it should be emphasized that because of its relatively young age, Monogem could conduce to the electron spectrum to higher energy than the other two SNRs. Moreover, Vela ($\tau=1.1\times10^4\,\text{yr}, R=0.29\,\text{kpc}$),  an even younger nearby source, might be able to extend the spectrum excess to TeV-PeV~\cite{23}.
Some recent studies~\cite{DiMauro:2020cbn,Evoli:2020szd,Fang:2020dmi} on the contribution of SNRs to primary electrons are consistent with the above analysis. What's more, utilizing the multi-messenger analysis method can constrain the injected electron spectrum properties of the SNRs (see Ref.~\cite{Manconi:2018azw,DiMauro:2017jpu} for details). It should be noted that some models suggest SNRs can also produce secondary positrons~\cite{Blasi:2009hv,Ahlers:2009ae,Mertsch:2014poa}, which is not considered in our study.

In 2021, the newly released electron/positron spectra of AMS-02 has reached around 1 TeV~\cite{13}, which provides an excellent opportunity for further study of the primary electron excess at high energies. Following the analysis in Ref.~\cite{li2015}, we focus on the flux difference between CR electron and positron (i.e., $\Delta \Phi = \Phi_{e^-}-\Phi_{e^+}$ whose error bar was obtained by applying the error transfer formula). In general, the flux of CR electrons contains three components: the primary electrons produced by supernova explosion (SNe), the secondary electrons from collisions of CRs with gases, and the extra component due to DM annihilation or pulsars. However, CR positrons contain only the secondary and the extra component. In the $\Delta \Phi$ data, the extra component of CR electrons and positrons can cancel out each other. For the secondary component, since the flux of secondary positrons is slightly larger than the secondary electrons, they only partially cancel out. Considering that the secondary component is much smaller than the primary one, the primary electron flux can be approximated as $\Delta \Phi = \Phi_{e^-}-\Phi_{e^+}$.

Using these data, we investigate the hardening behavior of primary electron spectrum with nearby SNRs model and calculate its CR proton fluxes which may correspond the spectral bump of protons around 10 TeV~\cite{Yue:2019sxt,Qiao:2019san,Fang:2020cru,Fornieri:2021sqq}. The CR proton spectrum we used here is measured by DAMPE~\cite{dampe-proton} which shows a strong evidence of the softening at $\rm \sim 10~TeV$ at a high confidence level.

This work is organized as follows. In Sec.~II, we briefly review the propagation model of CRs adopted in this work. Then in Sec.~III, we perform the fit analysis with the Markov Chain Monte Carlo (MCMC) method to investigate the contribution of nearby SNRs to the primary electron excess and proton spectral bump and yield the best fit parameters. Finally in Sec.~IV, we summarize our work.

\section{The cosmic rays propagation model}
\label{II}
Generally, the propagation of CRs can be described by~\cite{28}:
\begin{eqnarray}
\frac{\partial N}{\partial t}=Q+\nabla\cdot(D_{xx}\nabla N-\mathbf{V_c}N)+\frac{\partial}{\partial{p}}p^2D_{pp}\frac{\partial}{\partial{p}}\frac{N}{p^2}-\frac{\partial(N\dot{E})}{\partial{E}},
\label{eq1}
\end{eqnarray}
where $N$ is the number density of particles in the interval $d^3\mathbf{r}dE$, and $Q$ is the source distribution function including the CR particle generation, injection, fragmentation, decay, etc. $D_{xx}$ is the spatial diffusion coefficient, which is of the form $D_{xx}=\beta^\eta D_0(R/R_0)^\delta$, where $R$ and $\delta$ are the rigidity and diffusion index of the CR particle, and $\beta$ is the velocity in unit of light speed. The coefficient $\eta$ is the velocity-dependence term, which is equal to 1 in the traditional Diffusion-Reacceleration (DR) model~\cite{DiBernardo:2009ku}. $D_{pp}=p^2V^2_A/(9D_{xx})$ is the diffusion coefficient in momentum space, where $V_A$ is the $\rm{Alfv\Acute{e}n}$ velocity. $\mathbf{V_c}$ denotes the convection velocity and $\dot{E}={dE}/{dt}$ is the energy loss rate. Finally, the propagation zone of CRs is confined in a cylindrical slab with a half thickness $z_h$.

For CR positrons and electrons, it is essential to calculate the energy loss when propagation and it primarily arises in the processes of synchrotron radiation and inverse Compton scattering. In the Galactic magnetic field, the synchrotron radiation energy loss of the electron can be written as
\begin{eqnarray}
\frac{dE}{dt}=-\frac{4}{3}c\sigma_TU_B\gamma^2\beta^2 \sim -2.53\times10^{-18}(\frac{B}{\mu\rm G})^2(\frac{E}{\rm {GeV}})^2\frac{\rm{GeV}}{\rm{s}},
\label{eq2}
\end{eqnarray}
where $\sigma_T$ is the Thomson cross section, $U_B$ is the magnetic field energy density, and $B$ is the strength of the magnetic fields. For inverse Compton scattering, star lights and cosmic microwave background photons play the key role. The total energy loss rate can be described by
\begin{eqnarray}
b(E)=b_0(E)E^2.
\label{eq3}
\end{eqnarray}
Here $b_0(E)$ is the energy loss coefficient which is decided by synchrotron radiation and inverse Compton scattering.

The propagation equation can be solved semi-analytically with the help of Green’s function~\cite{Ginzburg,Strong}, which is expressed as
\begin{eqnarray}
G(r,t,E \gets r_0,t_0,E_0)=\delta(t-t_0-\tau)\frac{1}{b(E)(\pi\lambda^2)^{\frac{3}{2}}}\text{exp}[-\frac{(r-r_0)^2}{\lambda^2}],
\label{eq4}
\end{eqnarray}
where $\tau=\frac{1}{b_0}(\frac{1}{E}-\frac{1}{E_0})$ and $\lambda^2=4\int_{E}^{E_0}dE^\prime\frac{D(E^\prime)}{b(E^\prime)}$
is the diffusion distance of particles. Then, the corresponding solution can be achieved by
\begin{eqnarray}
N(r,t,E)=\iiint dr_0dt_0dE_0G(r,t,E \gets r_0,t_0,E_0)Q(r_0,t_0,E_0).
\label{eq7}
\end{eqnarray}

In this work, we consider SNR as a point source with the burst-like injection. In such a model, the source term can be described as
\begin{eqnarray}
Q(r,t,E)=Q(E)\delta(r-r_s)\delta(t-t_s),
\label{eq8}
\end{eqnarray}
where $r_s$ and $t_s$ are the location and injection time of the source, respectively. The solution of Eq.(\ref{eq1}) in this case can be directly derived with the following formula
\begin{eqnarray}
N(r,t,E)=\frac{b(E_s)}{b(E)(\pi\lambda^2)^{\frac{3}{2}}}\text{exp}[-\frac{(r-r_s)^2}{\lambda^2}]Q(E_s).
\label{eq9}
\end{eqnarray}

Charged CR particles can be accelerated in SNR to very high energy through a shock acceleration mechanism, and the energy spectrum of high energy particles produced by this mechanism can be modeled by a power-law function with an exponential cut-off as follows
\begin{eqnarray}
Q(E)=Q_0(\frac{E}{1\rm{GeV}})^{-\alpha}\text{exp}(-\frac{E}{E_c}),
\label{eq10}
\end{eqnarray}
where $Q_0$ denotes the normalization of injection spectrum. $\alpha$ and $E_c$ are the power-law spectral index and cut-off energy, respectively.

Due to the influence of the Solar wind plasma, known as the Solar modulation, low energy charged CR particles would be greatly suppressed when they break in the Solar system. By solving the equation of CR propagation in the Solar system, the approximate solution of CR particle flux $J$ after the Solar modulation is given as a function of the flux $J_{\text{LIS}}$ before the Solar modulation~\cite{Gleeson}:
\begin{eqnarray}
J(E)=\frac{E^2-m^2}{(E+|Z|\phi)^2-m^2}J_{\text{LIS}}(E+|Z|\phi),
\label{eq11}
\end{eqnarray}
where $E$, $m$ and $Z$ are the energy, mass and charge number of CR particles, respectively. $\phi$ is so-called solar modulation potential.

\section{The contribution of a single nearby SNR to the primary electron excess and proton spectral bump}

In general, it is hard to solve the CR propagation equation analytically, especially considering some complicated cases such as the distribution of magnetic field, interstellar medium and CR sources. Thus, we use the package \text{GALPROP}~\cite{29} to calculate the propagation equation numerically. In this work, we adopt the DR model and fix the CR propagation parameters to be the best-fit values obtained from the observational data of ${\rm B/C}$, ${\rm ^{10}B{\rm e}/^{9}B{\rm e}}$ and protons~\cite{Yuan:2018vgk}. The main propagation parameters for DR model are shown in TABLE~\ref{table_1}, where $E^{\rm proton}_{\rm br}$ and $\gamma_{1(2)}$ are the proton break rigidity and injection index, and the other parameters have been reported in the Sec.~\ref{II}.
\begin{table}[!htb]
\begin{tabular}{cccccccc}
\hline\hline
$z_{h}$ & $D_{0}$ & $\eta$ & $R_0$ & $\delta$ & $V_A$ & $E^{\rm proton}_{\rm br}$& $\gamma_1/\gamma_2$\\
$\rm (kpc)$ & ($10^{28}$ cm$^2$ s$^{-1}$) &  & $\rm (GV)$ &  & (km s$^{-1}$) & (GeV) &  \\
\hline
6.11&6.46&-0.48&4.0&0.41&29.4&10.35&2.028/2.405 \\
\hline\hline
\end{tabular}
\caption {The CR propagation parameters for DR model.}
\label{table_1}
\end{table}

Further, we use the package \text{CosmoMC}~\cite{32} as a MCMC engine to explore the parameter space of CR backgrounds and nearby sources, and find the best-fit parameters. The free parameters we used here to fit the latest AMS-02 $\Delta \Phi$ data are shown in TABLE \ref{Table_2}, where $\Gamma_1$, $\Gamma_2$ and $\rho_{br}$ are the injected spectral indexes and the break rigidity of the electron background, respectively. $N_e$ is the normalized flux of electron at 25 GeV and $\phi$ denotes solar modulation potential. The results of the best-fit parameters considering only the background for DR model are given in the last column of TABLE~\ref{Table_2}. We find a minimal $\chi^2/\rm d.o.f.$ of 6.01, which is too high to be unacceptable.
This further confirms not only the hardening behavior of primary electrons at high energies but also the necessity of extra CR electron sources from the upper left panel of FIG.~\ref{fig1} (the green dashed line). So we will investigate the possible contribution of SNRs as the extra excess of CR electrons.

\subsection{Explaining the AMS-02 primary electron data}

\begin{figure}[t]
\includegraphics[width=8cm,height=6.5cm]{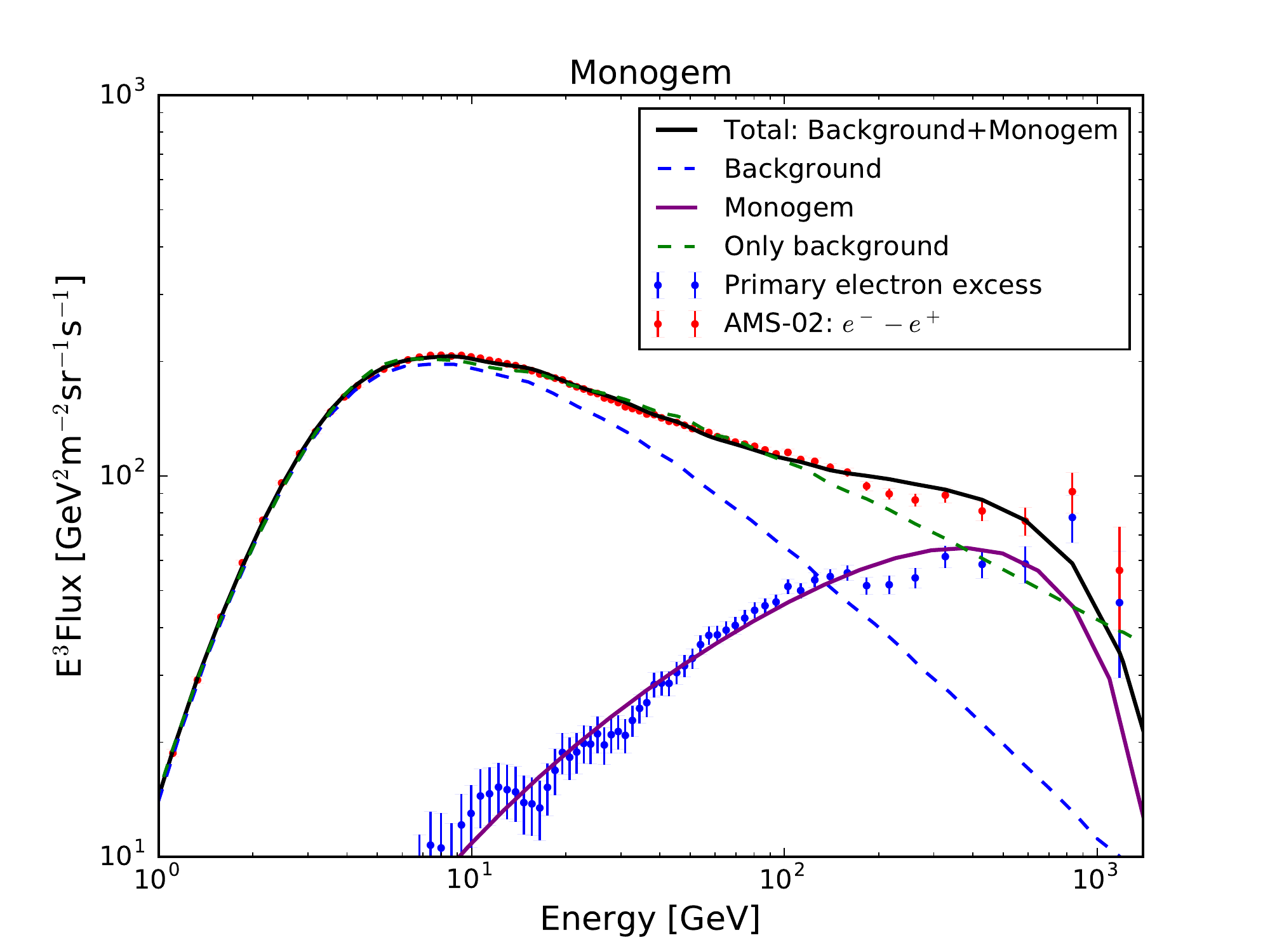}
\includegraphics[width=8cm,height=6.5cm]{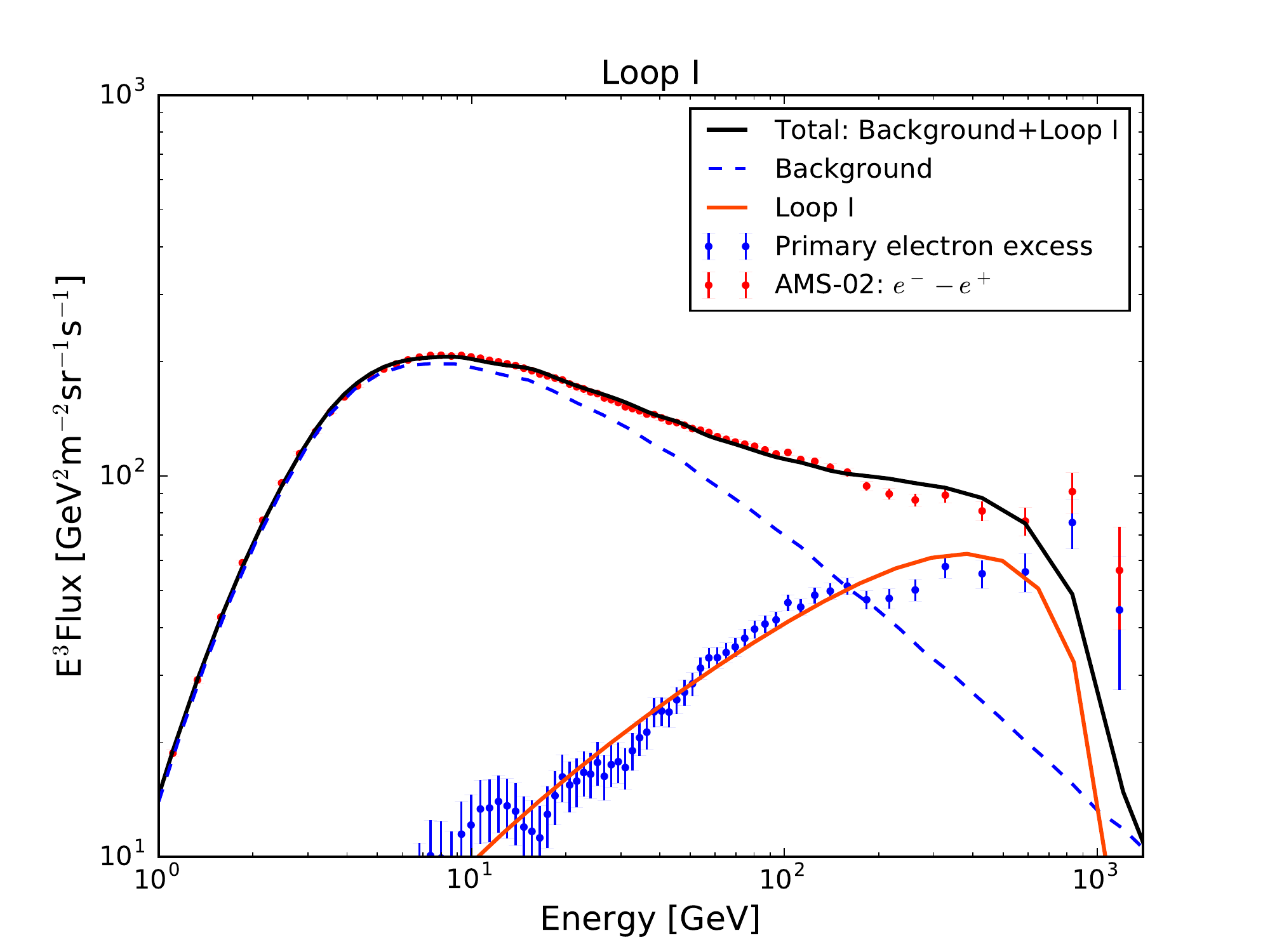}
\includegraphics[width=8cm,height=6.5cm]{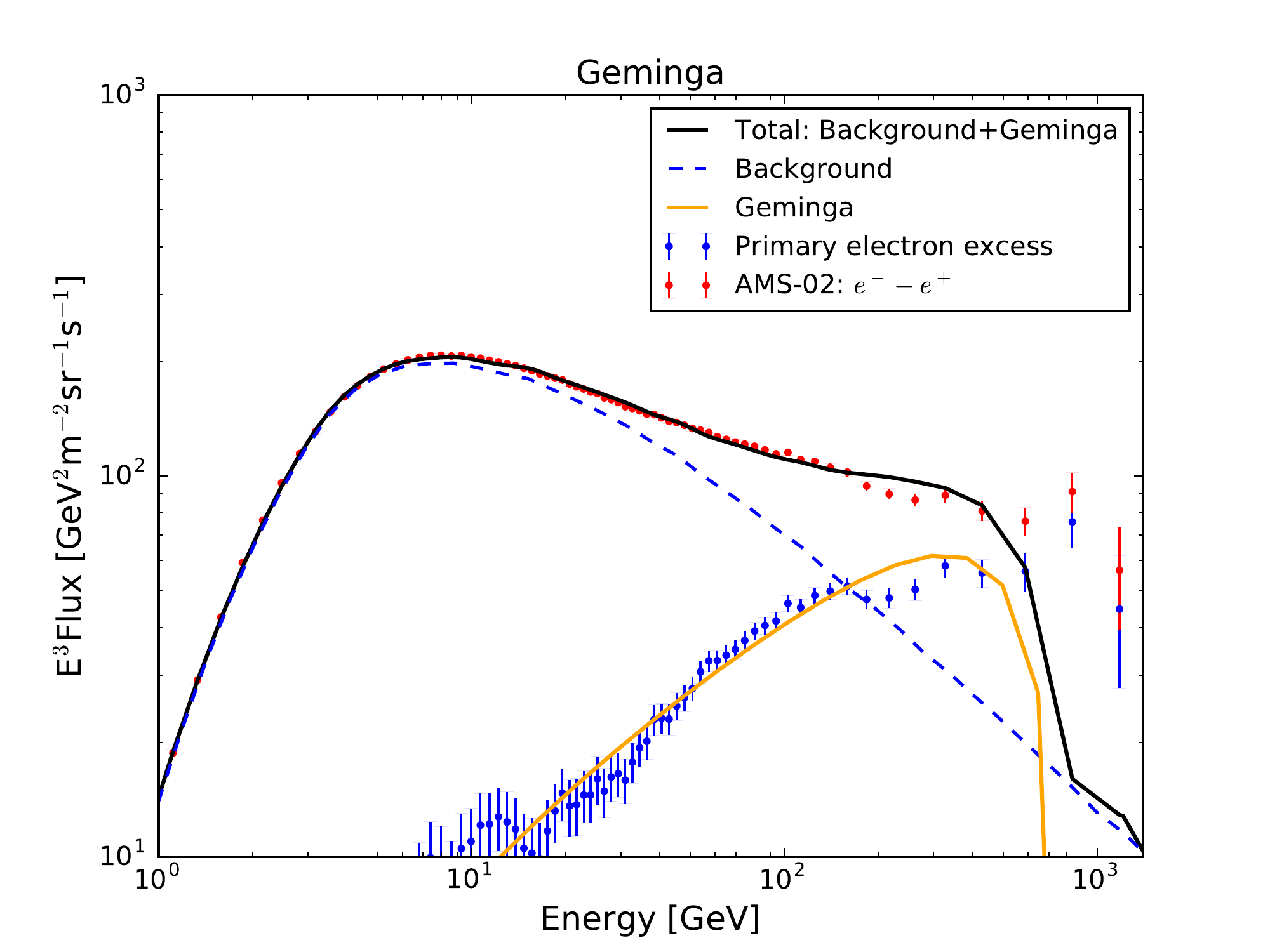}
\caption{The global best-fit results of three single-source models for AMS-02 primary electron spectrum. The blue dashed line represents the best-fit DR background for each single-source scenario and the color solid line is the corresponding SNR contribution. The global best-fit to $\Delta \Phi$ data is denoted by the black solid line. The green dashed line represents the DR background.}
\label{fig1}
\end{figure}

To investigate which SNR can provide a possible explanation of the primary electron excess, we separately calculate the contribution of the following three sources: Monogem, Loop I and Geminga. The reasons for choosing these three candidates have been discussed in the first section.
In specific calculation, the free parameters include the DR background parameters as well as the three electron injection spectrum parameters of SNR: the injection spectrum normalization ($Q_0$), power-law spectral index ($\alpha$) and cut-off energy ($E_c$). The specific best-fit values of these three single sources are summarized in TABLE~\ref{Table_2}. From the corresponding minimal $\chi^2/\rm d.o.f.$, we find that Monogem ($\chi^2/\rm d.o.f.$=1.28) scenario is superior to Loop I and Geminga for the primary electron excess, which
is also evident in FIG.~\ref{fig1}.

\begin{table}[h]
\begin{tabular}{p{5.5cm}p{2cm}p{2cm}p{2cm}p{2cm}}
\hline\hline
 Parameters & Monogem & Loop I & Geminga & DR \\
\hline
$\Gamma_1$ & $2.107^{+0.027}_{-0.035}$ & $1.901^{+0.022}_{-0.020}$ & $1.995^{+0.019}_{-0.014}$ & $1.302^{+0.025}_{-0.025}$ \\
$\rho_{br}$ ($\rm {GV}$) & $5.374^{+0.056}_{-0.057}$ & $5.318^{+0.040}_{-0.047}$ & $5.244^{+0.054}_{-0.055}$ & $5.417^{+0.025}_{-0.028}$\\
$\Gamma_2$ & $3.058^{+0.004}_{-0.003}$ & $3.015^{+0.006}_{-0.006}$ & $3.026^{+0.004}_{-0.004}$ & $2.720^{+0.003}_{-0.003}$\\
$N_e$ ($10^{-9}\rm {cm}^{-2}\rm {sr}^{-1}\rm s^{-1}\rm {MeV}^{-1}$) & $1.236^{+0.007}_{-0.007}$ & $1.245^{+0.010}_{-0.007}$ & $1.270^{+0.007}_{-0.008}$ & $1.236^{+0.007}_{-0.005}$ \\
$Q_0$ ($10^{50}\rm {GeV}^{-1}$)  & $9.280^{+0.522}_{-0.561}$ & $6.899^{+0.673}_{-0.705}$ & $13.024^{+0.093}_{-0.086}$  & -- \\
$\alpha$ & $2.127^{+0.013}_{-0.013}$ & $1.964^{+0.021}_{-0.024}$ & $1.925^{+0.012}_{-0.013}$ &-- \\
$E_c$ ($\rm {TeV}$) & $1.165^{+0.091}_{-0.064}$ & $1.321^{+0.118}_{-0.104}$ & $1.496^{+0.071}_{-0.059}$ &-- \\
$\phi$ ($\rm {GV}$) & $1.245^{+0.007}_{-0.006}$ & $1.182^{+0.013}_{-0.013}$ & $1.227^{+0.010}_{-0.010}$ &$0.626^{+0.019}_{-0.017}$\\
$\chi^2/\rm {d.o.f}$ & 1.28 & 1.73 & 2.28 &6.01\\
\hline\hline
\end{tabular}
\caption{The best-fit parameters of the latest AMS-02 $\Delta\Phi$ data for the DR background and single-source models.}
\label{Table_2}
\end{table}

\begin{figure}[h]
\includegraphics[width=12cm,height=8cm]{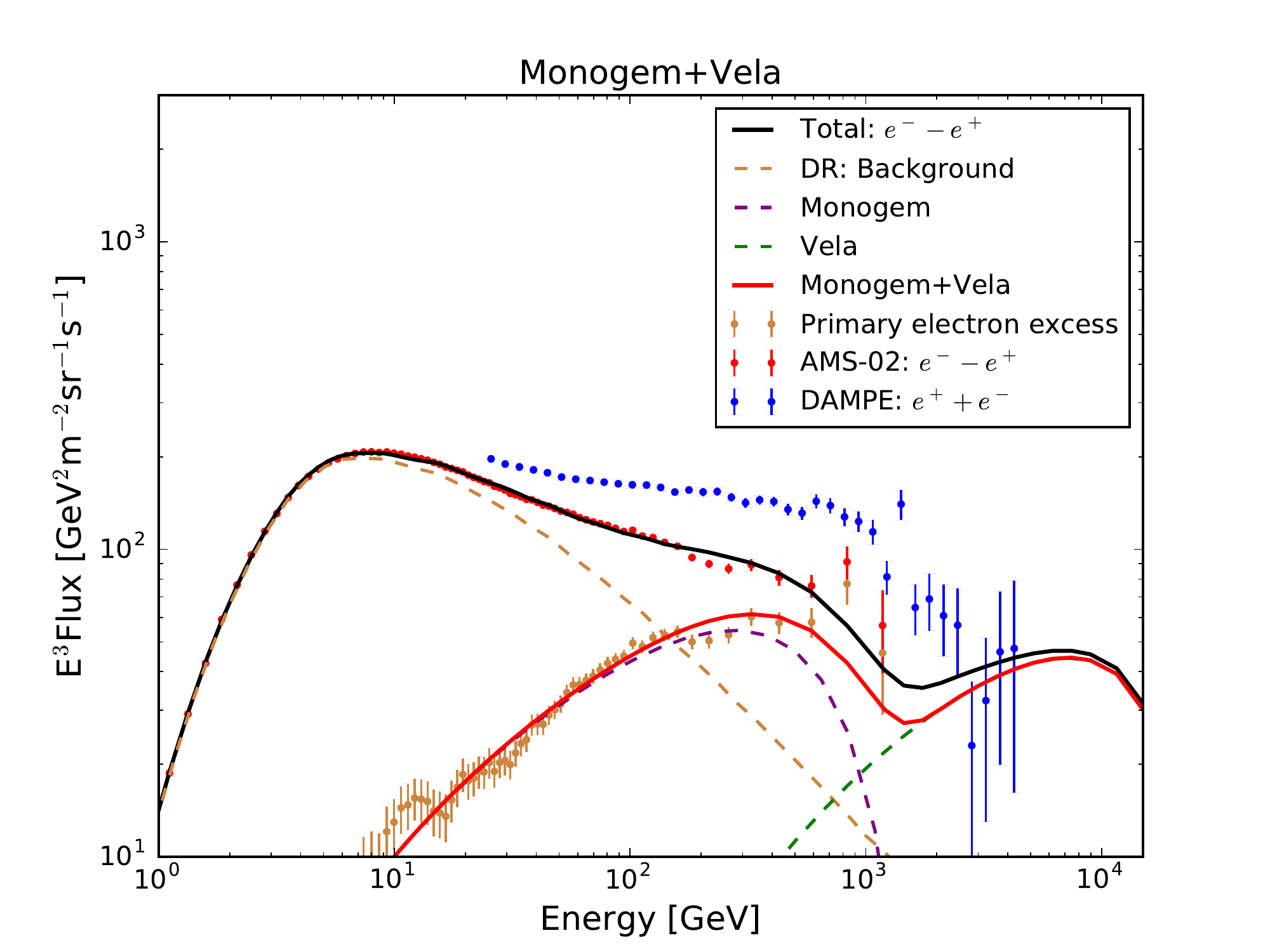}
\caption{Constraining the possible contribution of the Vela SNR on the generation of TeV CR electrons.}
\label{fig1.1}
\end{figure}

According to FIG.~\ref{fig1}, we can see that the contribution of Loop I and Geminga to electrons is mainly at hundreds of GeV, but Monogem can extend the electron spectrum to TeV region, which is the reason why Monogem has a better fitting result than the other two sources. As we know, Vela may dominate the CR electrons above 1 TeV because of its younger age and appropriate distance. Moreover, due to its small contribution to the low-energy electrons, Vela is not considered as a suitable candidate for the primary sub-TeV electron excess. Here,
we roughly predict the primary CR electron spectrum of Vela and use the $e^++e^-$ spectrum released by DAMPE~\cite{33} to give an upper limit. The corresponding result is shown in FIG.~\ref{fig1.1}, which implies that the primary electron spectrum may harden again at a few TeVs. DAMPE aims to measure the high-precision electron spectrum with energy range from $\sim$5 GeV to $\sim$10 TeV. With a large acceptance of $\sim 0.3 ~{\rm m^2~Sr}$,  it is expected to robustly probe
this possible spectral feature in the future.

\subsection{Explaining the DAMPE proton data}
Several earlier experiments had identified the proton spectrum hardening at about 300 GeV~\cite{atic-proton,cream-proton,pamela-proton,ams02-proton}. However, as measurements of the proton spectrum are extended to higher energies, a softening signature around 10 TeV has also been reported~\cite{34,35}, and this spectral feature was confirmed by DAMPE with high significance in 2019~\cite{dampe-proton}. Moreover, it should be pointed out that DAMPE is the only experiment so far that can cover both the spectral hardening and softening. We know that a single nearby SNR may naturally contribute a bump to the spectrum in the high energy range. Therefore, we expect to simultaneously explain the proton spectrum of DAMPE with a discrete nearby SNR candidate that can explain the primary electron excess.
\begin{table}[h]
\begin{tabular}{p{1.8cm}ccccccc}
\hline\hline
Sources & $E^{\rm{proton}}_{\rm{br}}$($\rm {GeV}$) & $\gamma_1/\gamma_2$ & $N_p^*$ & $Q^{\dagger}_0$ & $\alpha$ & $E_c$($\rm{TeV}$)&$\chi^2/\rm {d.o.f}$ \\
\hline
Monogem& $10.54^{+1.02}_{-0.72}$&$1.95^{+0.14}_{-0.14}$/$2.53^{+0.06}_{-0.05}$ &$3.29^{+0.18}_{-0.17}$& $9.55^{+1.79}_{-1.72}$ &$1.97^{+0.02}_{-0.02}$ &$88.08^{+19.29}_{-16.91}$&0.32  \\
Loop I& $10.47^{+1.76}_{-1.16}$&$2.08^{+0.23}_{-0.22}$/$2.46^{+0.08}_{-0.06}$ &$3.67^{+0.30}_{-0.43}$ &$7.32^{+1.98}_{-1.79}$ &$1.86^{+0.05}_{-0.04}$ &$70.89^{+12.56}_{-10.62}$&0.35  \\
\hline\hline
\end{tabular}\\
  $^*$ $N_p$ is in unit of $10^{-9}{\rm cm^{-2}~sr^{-1}~s^{-1}~MeV^{-1}}$;\quad
  $\dagger$ $Q_0$ is in unit of $10^{52}\rm {GeV}^{-1}$.
\caption{The best-fit parameters of the DAMPE proton data for the single-source models.}
\label{table_4}
\end{table}

For simplicity, we assume that a power-law background and a nearby SNR contribute to the observed CR proton fluxes. In this scenario, we separately calculate the contributions of Monogem and Loop I, which are the two candidates examined in the modeling of primary electron excess. We use the same MCMC method to fit the proton data from DAMPE. The best-fit values of free parameters are summarized in TABLE.~\ref{table_4}, where $E^{\rm proton}_{\rm br}$ is the break rigidity of the background and $N_p$ is the normalized proton flux at 100 GeV, and $Q_0$, $\alpha$ and $E_c$ are parameters of SNRs, respectively.
\begin{figure}[t]
\includegraphics[width=8cm,height=6.5cm]{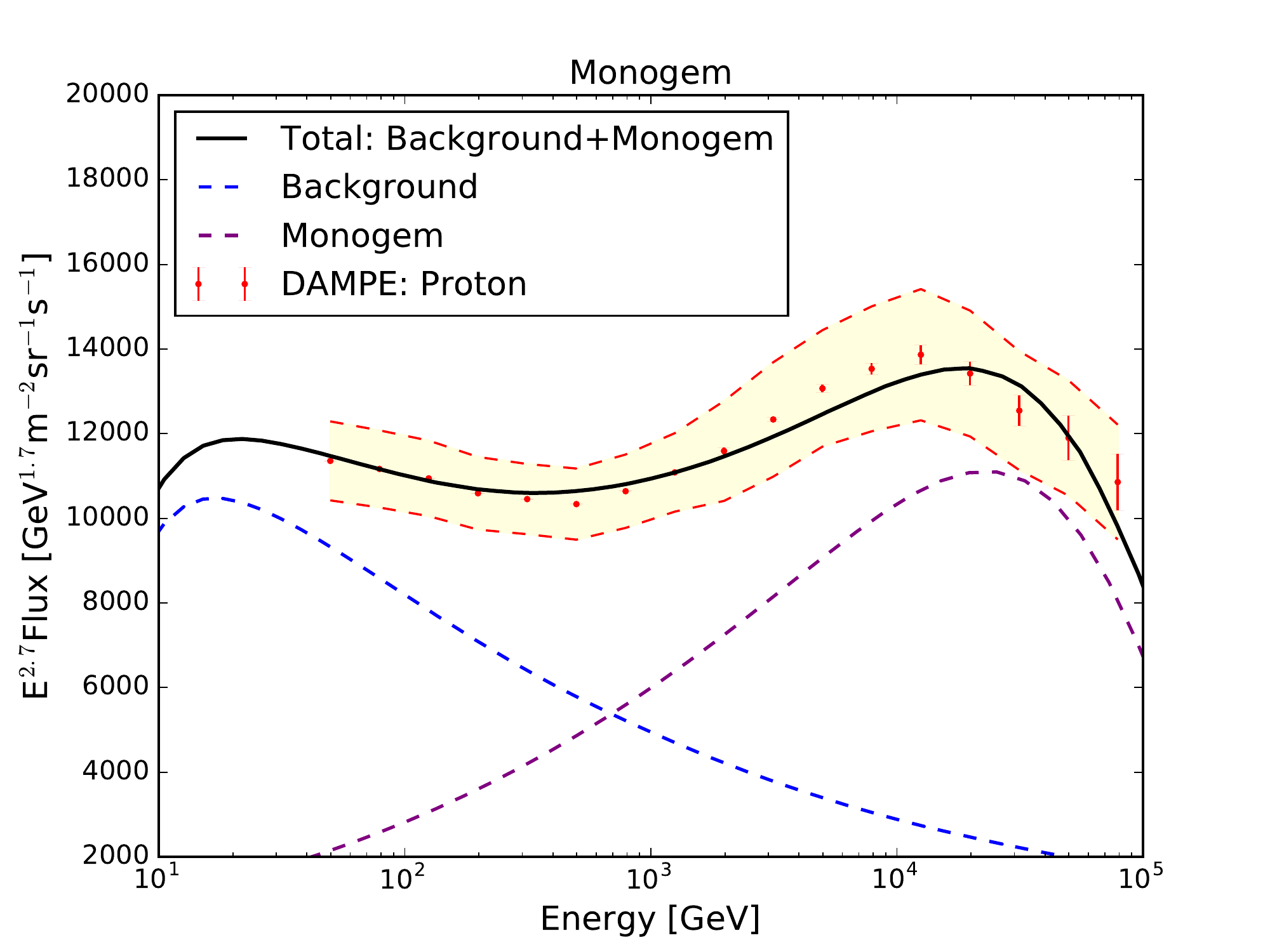}
\includegraphics[width=8cm,height=6.5cm]{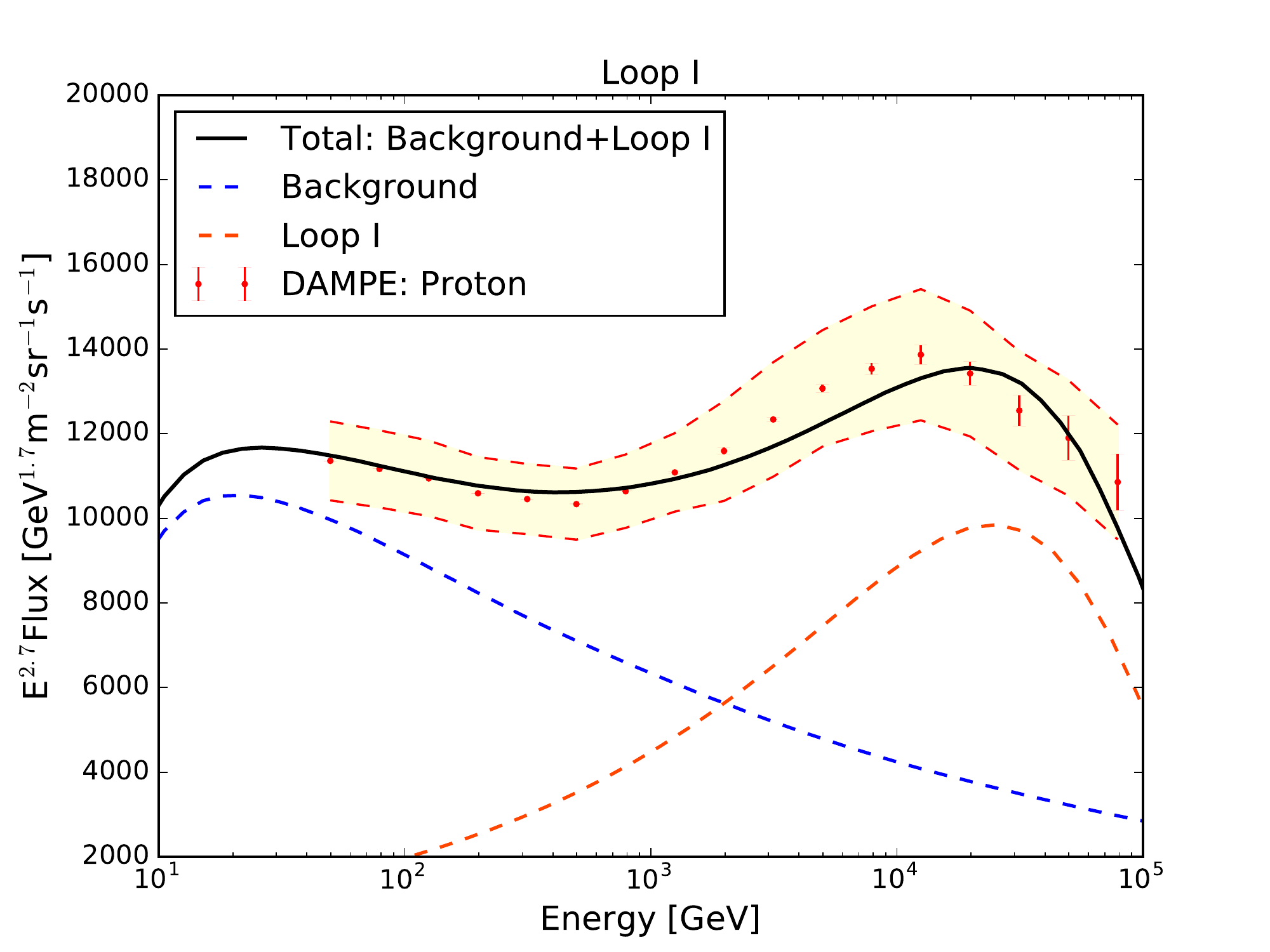}
\caption{The global best-fit results of Monogem (left) and Loop I (right) for DAMPE proton spectrum.}
\label{fig4}
\end{figure}

The corresponding best-fit results of DAMPE proton spectrum are shown in FIG.~\ref{fig4}. Obviously, both nearby single-source models can fit the proton data well. In addition, the conclusion of Ref.~\cite{36} shows that a nearby source located at a proper direction together with the background components can explain both the feature of proton spectrum and the amplitudes and phases of CRs anisotropy. So our results suggest that Monogem and Loop I may be appropriate candidates.

\section{CONCLUSION}
In this work, we simultaneously investigate the contribution of nearby SNRs to the primary electron excess at high energies and the proton spectral bump around 10 TeV. For the primary electron excess, we use the newly released electron/positron data from AMS-02 Collaboration, and focus on the difference between electron and positron (i.e., $\Delta \Phi = \Phi_{e^-}-\Phi_{e^+}$). After fitting with only the DR background model, we further prove the hardening behavior of primary electron spectrum. Then we discuss the possibility of the single-source model to account for this excess. The result indicates that Monogem is more favored by data than Loop I and Geminga, because the Monogem’s contribution to the electron spectrum can extend to $\sim 1~{\rm TeV}$. We also briefly discuss the possible contribution of Vela to the electron spectrum and use the DAMPE $e^++e^-$ data to check the rationality of our result. We find that the CR electron spectrum might harden again at a few TeVs, which could be probed by DAMPE in the near future. Since the DAMPE proton data covers both the spectral hardening and softening, we expect to fit the proton spectrum of DAMPE simultaneously with the nearby SNR that can explain the primary electron excess. Finally, we find that both the primary electron excess and the proton spectral bump may be mainly generated by Monogem.

\section*{Acknowledgement}
This work is supported by the National
Natural Science Foundation of China (Grants No. U1738210, No. U1738206, No. 12003069, No. 12047560, No. 12003074, No. 11773075, and No. U1738136), the 100
Talents Program of Chinese Academy of Sciences and the Entrepreneurship and Innovation Program of Jiangsu Province.

\appendix
\section{The combination of two SNRs model for fitting $\Delta\Phi$ data}
The single-source model has done a good job of explaining the excess of primary electrons. However, if we combine the contributions of possible sources, this should give better fitting results. For simplicity, we focus here only on some combinations of two middle-age nearby sources. This scenario includes three cases: Monogem + Loop I, Monogem + Geminga and Loop I + Geminga. Since the contribution of Monogem to the electron spectrum can extend to the sub-TeV region, we speculate that the Monogem + Loop I/Geminga cases would yield the best-fit results.

\begin{table}[t]
\centering
\begin{tabular}{p{3cm}p{4.3cm}p{4.3cm}p{4.3cm}}
\hline\hline
 & Monogem &Monogem &Loop I \\
Parameters &+&+&+\\
 &Loop I&Geminga&Geminga\\
\hline
$\Gamma_1$ & $2.104^{+0.016}_{-0.013}$ & $2.113^{+0.030}_{-0.032}$ & $2.109^{+0.030}_{-0.030}$  \\
$\rho_{br}$ ($\rm {GV}$) & $5.348^{+0.065}_{-0.061}$ & $5.387^{+0.062}_{-0.074}$ & $5.342^{+0.075}_{-0.076}$  \\
$\Gamma_2$ & $3.056^{+0.005}_{-0.005}$ & $3.048^{+0.004}_{-0.004}$ & $3.061^{+0.004}_{-0.004}$  \\
$N_e^*$  & $1.247^{+0.007}_{-0.007}$ & $1.270^{+0.006}_{-0.006}$ &  $1.234^{+0.008}_{-0.008}$   \\
$Q_0$ ($10^{50}\rm {GeV}^{-1}$)  & $1.102^{+0.033}_{-0.034}$/$4.877^{+0.568}_{-0.501}$ & $ 1.286^{+0.170}_{-0.157}$/$6.311^{+0.320}_{-0.364}$ & $2.994^{+0.176}_{-0.187}$/$8.820^{+0.422}_{-0.584}$   \\
$\alpha$ & $1.893^{+0.010}_{-0.009}$/$1.908^{+0.029}_{-0.025}$ & $1.900^{+0.016}_{-0.020}$/$1.813^{+0.014}_{-0.011}$ & $1.900^{+0.011}_{-0.011}$/$1.909^{+0.013}_{-0.012}$   \\
$E_c$ ($\rm {TeV}$) & $2.554^{+0.128}_{-0.094}$/$0.319^{+0.009}_{-0.009}$ & $2.399^{+0.182}_{-0.166}$/$0.316^{+0.008}_{-0.007}$ &$2.850^{+0.236}_{-0.232}$/$0.324^{+0.016}_{-0.014}$  \\
$\phi$ ($\rm {GV}$) & $1.251^{+0.007}_{-0.007}$ & $1.250^{+0.010}_{-0.009}$ & $1.244^{+0.012}_{-0.013}$  \\
$\chi^2/\rm {d.o.f}$ & 1.03 & 1.05 & 1.19  \\
\hline\hline
\end{tabular}\\
$^*$ $N_e$ is in unit of $10^{-9}{\rm cm^{-2}~sr^{-1}~s^{-1}~MeV^{-1}}$.
\caption{The best-fit parameters of the latest AMS-02 $\Delta\Phi$ data for the two SNRs model.}
\label{Table_3}
\end{table}

\begin{figure}[t]
\includegraphics[width=8cm,height=6.5cm]{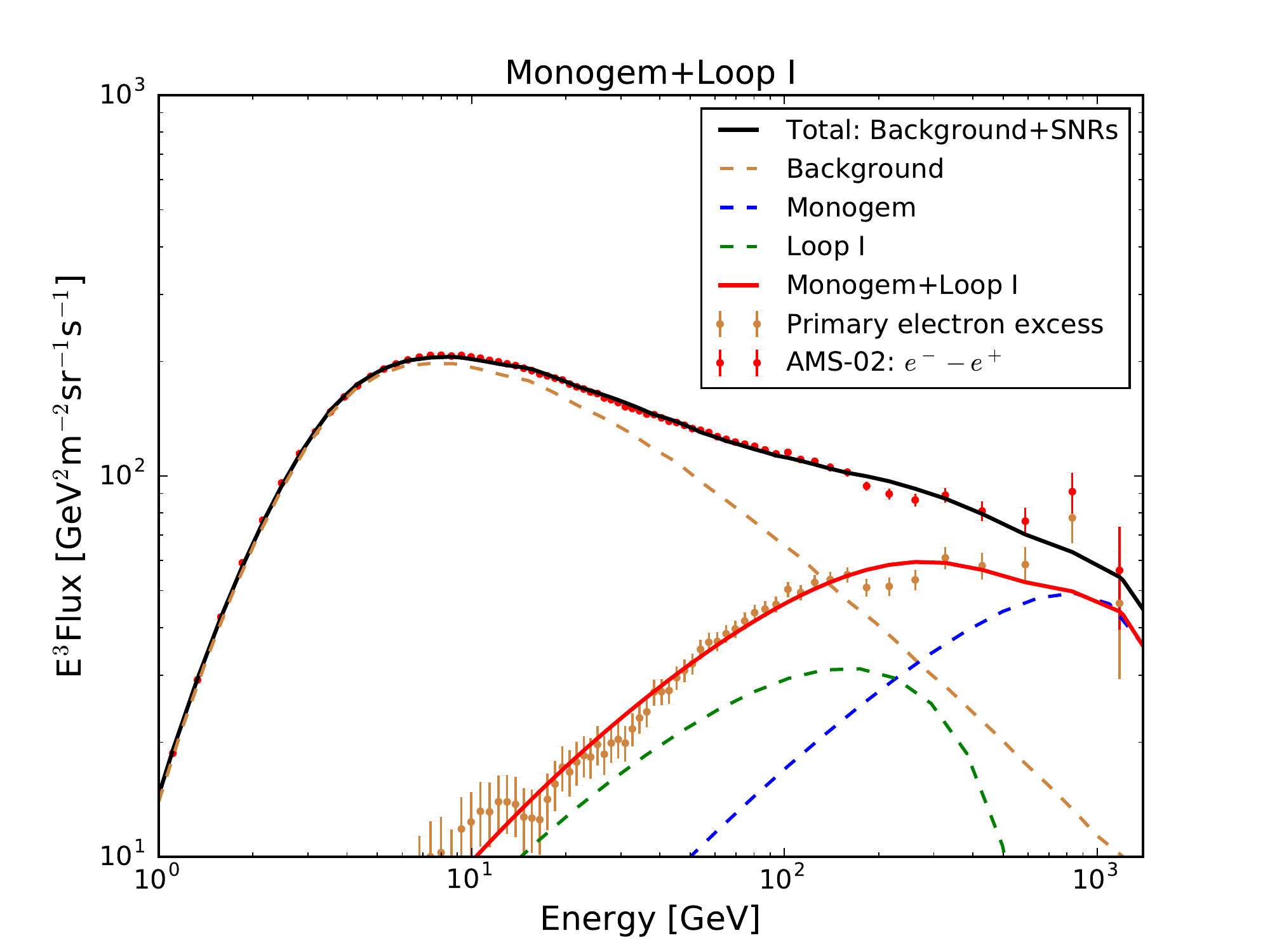}
\includegraphics[width=8cm,height=6.5cm]{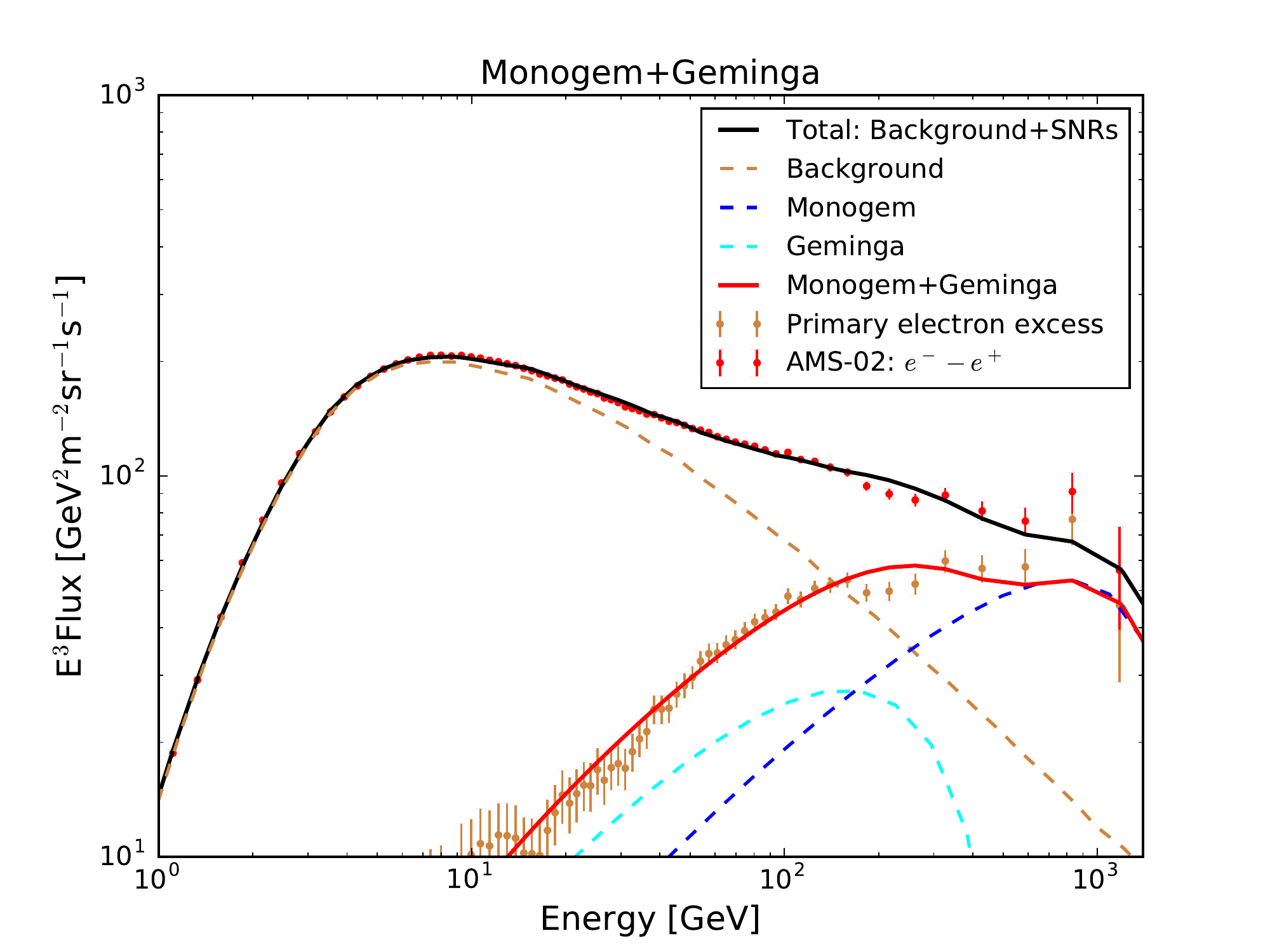}
\includegraphics[width=8cm,height=6.5cm]{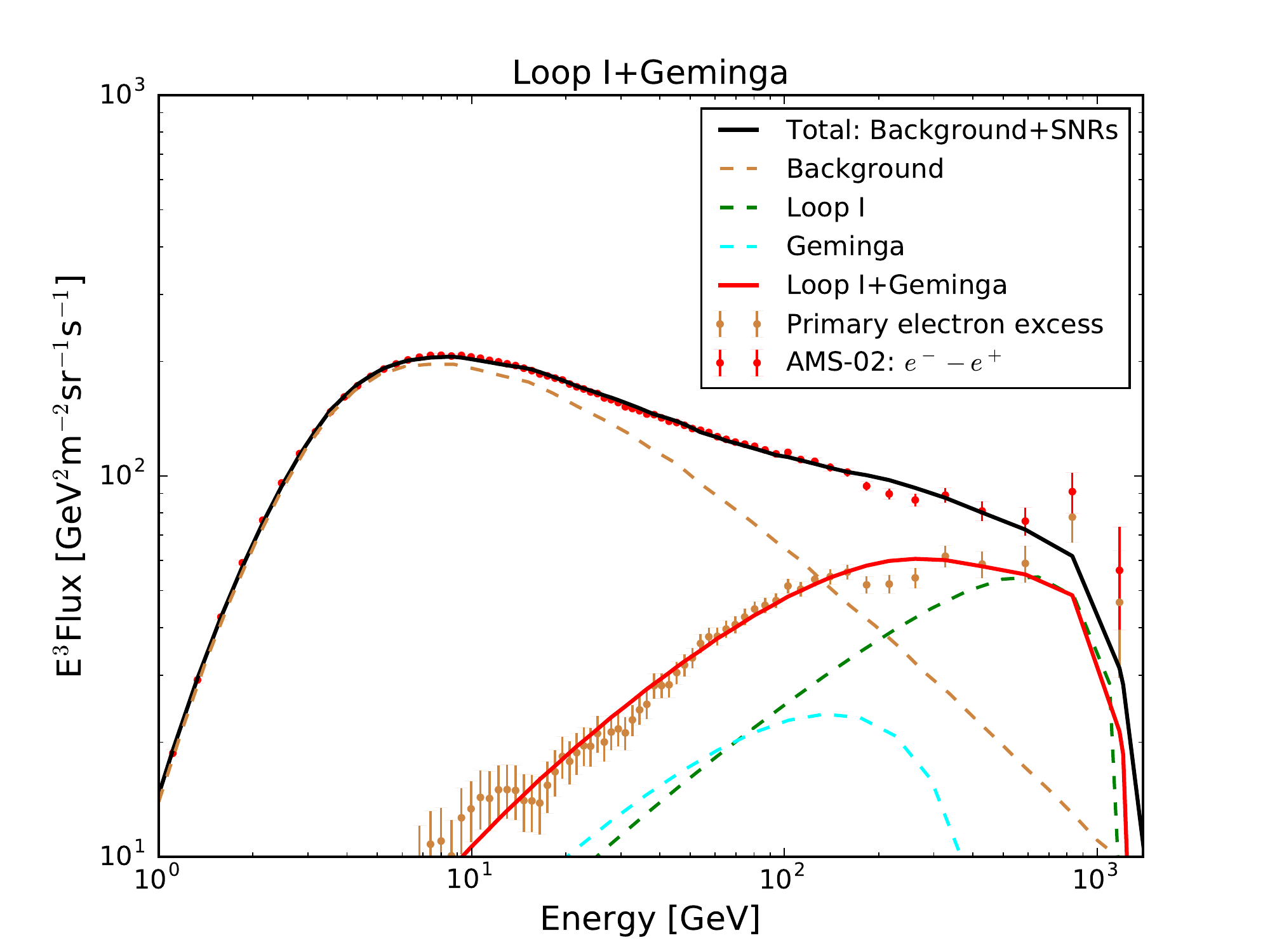}
\caption{Same as FIG.~\ref{fig1}, but for the two SNRs model.}
\label{fig2}
\end{figure}

FIG.~\ref{fig2} shows the global best-fit of the primary electron spectrum where we still set the SNR's parameters $Q_0$, $\alpha$ and $E_c$  free. The corresponding best-fit values are listed in TABLE.~\ref{Table_3}. We find that Monogem + Loop I/Geminga models both give the appropriate minimal $\chi^2/\rm d.o.f.$, where the minimum value is 1.03. However, the fitting of Loop I + Geminga model is not good enough, as neither of these two sources have larger enough contributions above 1 TeV.

\end{document}